\documentclass[11pt]{article}
\usepackage{graphicx} 
\usepackage{amsmath}
\usepackage{amsfonts}   
\usepackage{amssymb}

\begin{document}
\date{}
\title{{\bf{\Large Noncommutative Schwarzschild Black Hole and Area Law }}}
\author{
 {\bf {\normalsize Rabin Banerjee}$
$\thanks{E-mail: rabin@bose.res.in}},\, 
 {\bf {\normalsize Bibhas Ranjan Majhi}$
$\thanks{E-mail: bibhas@bose.res.in}} {\bf {\normalsize and}} 
{\bf {\normalsize Sujoy Kumar Modak}
\thanks{E-mail: sujoy@bose.res.in}}\\
 {\normalsize S.~N.~Bose National Centre for Basic Sciences,}
\\{\normalsize JD Block, Sector III, Salt Lake, Kolkata-700098, India}
\\[0.3cm]
}

\maketitle

{\bf Abstract:}\\
     Using a graphical analysis, we show that for the horizon radius $r_h\gtrsim 4.8\sqrt\theta$, the standard semiclassical Bekenstein-Hawking area law for noncommutative Schwarzschild black hole exactly holds for all orders of $\theta$. We also give the corrections to the area law to get the exact nature of the Bekenstein-Hawking entropy when $r_h<4.8\sqrt\theta$ till the extremal point $r_h=3.0\sqrt{\theta}$.\\\\\\

             The familiar notion of black holes as objects from which nothing, not even light, can escape was changed radically by the startling observation of Hawking \cite{Hawking1,Hawking2}. His analysis, based on quantum field theory in a curved background, revealed that black holes emit a spectrum that is analogous to thermal black body spectrum. It gives the black holes one of its thermodynamic properties making it consistent with the rest of physics.

     Before this, Bekenstein \cite{Beken1,Beken2,Beken3,Beken4} proposed that a black hole has an entropy $S_{{\textrm {bh}}}$ which is some finite multiple $\eta$ of its area $A$. He was not able to determine the exact value of $\eta$, but gave heuristic arguments for conjecturing that it was $\frac{ln 2}{8\pi}$. However, the first law of black hole mechanics imply that the black hole would have a temperature $T_h$ which is proportional to the surface gravity $\kappa$ of the black hole. Therefore from Bekenstein's argument and the first law of black hole mechanics one might say $T_h=\epsilon\kappa$ and $S_{{\textrm {bh}}}=\eta A$ with $8\pi\eta\epsilon=1$. Bekenstein proposed that $\eta$ is finite and it is equal to $\frac{ln 2}{8\pi}$. Then one would get $\epsilon=\frac{1}{ln 2}$ and so $T_h=\frac{\kappa}{ln 2}$. Later on Hawking realised that Bekenstein's idea was consistent. In fact, he found that the black hole temperature is $T_h=\frac{\kappa}{2\pi}$, so that $\epsilon=\frac{1}{2\pi}$ and hence $\eta=\frac{1}{4}$. This leads to the famous Bekenstein-Hawking area law for entropy of black hole $S_{{\textrm {bh}}}=\frac{A}{4}$.

        All these calculations were based on the semiclassical concept and also on a commutative spacetime. The standard Bekenstein-Hawking area law is known to get corrections, either due to quantum geometry \cite{Kaul,Amit,Page} or back reaction effects \cite{Fursaev,Majhi2}. We are interested in obtaining the modifications to the area law due to noncommutative spacetime. Noncommutativity is expected to be relevant at the Planck scale where it is known that usual semiclassical considerations break down. It is therefore reasonable to expect that noncommutativity would modify the standard area law in a nontrivial manner. We show that this is indeed true. Although there has been some analysis \cite{Spall,Majhi1} in this direction, these are mostly qualitative and incomplete.

      In this paper we will consider the noncommutative Schwarzschild metric. This is obtained by solving the Einstein equation with a matter source as a Gaussian mass density smeared by the noncommutative paramater $\theta$ \cite{Smail,Miele,Park,Majhi1,Nozari}. The metric represents an anisotropic fluid type matter instead of the usual isotropic one. But in the limit $r<<\sqrt\theta$ and $r>>\sqrt\theta$ the metric again regains the isotropic nature. We will show that inside the event horizon, for small values of $r$ ($r<<\sqrt\theta$), the metric reduces to the de-Sitter metric with constant positive scalar curvature. Therefore, in contrast to the gravitational collapse of matter to a point, one finds a de-Sitter core surrounding the close vicinity of the usual singularity $r=0$ which is in agreement with earlier works \cite{Frolov,Dym}. Also, outside the horizon in the limit $r>>\sqrt\theta$ the standard Schwarzschild vacuum solution is reproduced. In this sense a singularity free black hole solution is obtained.

     The first step to study the noncommutative effects on the usual area law is to obtain the Hawking temperature. As shown in a paper involving two of us \cite{Majhi1}, the connection of this temperature with the surface gravity ($\kappa$), even for the noncommutative case, has the same functional form $T_h=\frac{\kappa}{2\pi}$ as in the commutative picture. The Hawking temperature, in a closed form, follows from this relation. We find that the physically well defined region of the noncommutative black hole is defined by $r_h\geq 3.0\sqrt\theta$, where $r_h$ is the horizon radius.

      Next, using the Gibbs form of first law of thermodynamics, the entropy is computed. We show analytically in the leading order in $\theta$, in the regime $\frac{r^2_h}{4\theta}>>1$, that the area law is just a noncommutative deformation of the usual semiclassical area law. Since higher order analytical computations are technically very involved we take recourse to a graphical analysis. We plot $\frac{dS_{{\textrm{bh}}}}{dr_h}$, where $S_{{\textrm{bh}}}$ is the Bekenstein-Hawking entropy, as a function of the horizon radius ($r_h$). This shows that when $r_h\geq4.8\sqrt\theta$ the noncommutative version of area law holds good to all orders in $\theta$. However when $r_h<4.8\sqrt\theta$ there is a deviation from this area law. We then find out the corrections so that the modified area law gives the nature of Bekenstein-Hawking entropy of the noncommutative Schwarzschild black hole in the physically defined region ($r_h\geq3.0\sqrt\theta$). We also observe that these corrections involve exponentials of the noncommutative horizon area as well as error functions.


        The fact is that ``gravitation'' is a manifestation of the ``curvature'' of spacetime and the presence of the gravitating objects are responsible for this curvature. Therefore, inclusion of noncommutative effect in gravity can be done in two ways. Directly take the spacetime as noncommutative, $[x_\mu,x_\nu]=i\theta_{\mu\nu}$ and use the Seibarg-Witten map to recast the gravitational theory (in noncommutative space) in terms of the corresponding theory in usual (commutative space) variables. This leads to correction terms (involving powers of $\theta{\mu\nu}$) in the various expressions like the metric, Riemann tensor etc. This approach has been adopted in \cite{Obre,Chai,Saha,Kobak,Samanta}. Alternatively,  incorporate the effect of noncommutativity in the mass term of the gravitating object. Here the mass density, instead of being represented by a Dirac delta function, is replaced by a Gaussian distribution. This approach has been adopted in \cite{Spall,Majhi1,Smail,Miele,Park}. The two ways of incorporating noncommutative effects in gravity are, in general, not equivalent. Here we follow the second approach, furthering our investigation \cite{Majhi1} on the computation of thermodynamic entities and area law for the noncommutative Schwarzschild black hole.

     The usual definition of mass density in terms of the Dirac delta function in commutative space does not hold good in noncommutative space because of the position-position uncertainty relation. In noncommutative space mass density is defined by replacing the Dirac delta function by a Gaussian distribution of minimal width $\sqrt\theta$  in the following way \cite{Smail}
\begin{eqnarray}    
\rho_{\theta}(r) = \dfrac{M}{{(4\pi\theta)}^{3/2}}e^{-{\frac{r^2}{4\theta}}}
\label{1.01}
\end{eqnarray}
where the noncommutative parameter $\theta$ is a small ($\sim$ Plank length$^2$) positive number. Using this expression one can write the mass of the black hole of radius $r$ in the following way
\begin{eqnarray}
m_\theta(r) = \int_0^r{4\pi r'^{2}\rho_{\theta}(r')dr' }=\frac{2M}{\sqrt\pi}\gamma(3/2 , r^2/4\theta) 
\label{1.02}
\end{eqnarray}
where $\gamma(3/2 , r^2/4\theta)$ is the lower incomplete gamma function defined as
\begin{eqnarray}
\gamma(a,x)=\int_0^x t^{a-1} e^{-t} dt.
\label{1.021}
\end{eqnarray}
In the limit $\theta\rightarrow 0$ it becomes the usual gamma function $(\Gamma_{{\textrm{total}}})$. Therefore $m_\theta(r)\rightarrow M$ is the commutative limit of the noncommutative mass $m_\theta(r)$.

   To find a solution of Einstein equation with the noncommutative mass density of the type (\ref{1.01}), the temporal component of the energy momentum tensor ${(T_{\theta})}_\mu^\nu$ is identified as, ${(T_{\theta})}_t^t=-\rho_\theta$. Now demanding the condition on the metric coefficients ${(g_\theta)}_{tt}=-{(g_\theta)}^{rr}$ for the noncommutative Schwarzschild metric and using the covariant conservation of energy momentum tensor ${(T_{\theta})}_\mu^\nu~_{;\nu}=0$, the energy momentum tensor can be fixed to the form,
\begin{eqnarray}
{(T_\theta)}_\mu^\nu={\textrm {diag}}{[-\rho_\theta,p_r,p',p']},
\label{1.022}
\end{eqnarray}
 where, $p_r=-\rho_\theta$ and $p'=p_r-\frac{r}{2}\partial_r\rho_\theta$. This form of energy momentum tensor is different from the perfect fluid because here $p_r$ and $p'$ are not same,
\begin{eqnarray}
p'=\Big[\frac{r^2}{4\theta}-1\Big]\frac{M}{(4\pi\theta)^{\frac{3}{2}}}e^{-\frac{r^2}{4\theta}}
\label{ref2}
\end{eqnarray}
i.e. the pressure is anisotopic. But for $r<<\sqrt\theta$, the first term in (\ref{ref2}) drops out and $p'=-\rho_\theta=p_r$, i.e. the energy-momentum tensor takes the isotropic form. When $r\rightarrow 0$ the energy density tends to a constant value $-\frac{M}{(4\pi\theta)^{\frac{3}{2}}}$. On the other hand, at the large values of $r$ ($r>>\sqrt\theta$) all the components of the energy-momentum tensor very quickly tend to zero and so the pressure is again isotropic and the Schwarzschild vacuum solution is well applicable. 

     The solution of Einstein equation (in $c=G=1$ unit) ${(G_\theta)}^{\mu\nu}=8\pi{{(T_\theta)}^{\mu\nu}}$, using (\ref{1.022}) as the matter source, is given by the line element \cite{Smail},   
\begin{eqnarray}
ds^2=-f_\theta(r)dt^2+\frac{dr^2}{f_\theta(r)}+r^2d\Omega^2;\,\,\, f_\theta(r)=-{(g_\theta)}_{tt}=\left(1-\frac{4M}{r\sqrt\pi}\gamma(\frac{3}{2},\frac{r^2}{4\theta})\right)
\label{1.04}
\end{eqnarray}
 Incidentally, this is same if one just replaces the mass term in the usual commutative Schwarzschild space-time by the noncommutative mass $m_\theta(r)$ from (\ref{1.02}). Also observe that for $r>>\sqrt\theta$ the above noncommutative metric reduces to the standard Schwarzschild form.

     The metric (\ref{1.04}) represents a self-gravitating, anisotropic fluid type matter. The existence of the radial pressure in the small length scale ($r<<\sqrt\theta$) is due to the quantum vacuum fluctuation and it balances the inward gravitational pull to prevent the collapse of the matter to a point. This is reminiscent of earlier works \cite{Frolov,Dym} where such a phenomenon is associated with the occurence of a de-Sitter metric inside the black hole $(f_\theta(r)<0)$. As we now show the introduction of noncommutativity naturally induces a de-Sitter metric for $r<<\sqrt\theta$. In this limit the metric coefficient $f_\theta(r)$ in (\ref{1.04}) reduces to,
\begin{eqnarray}
f_\theta(r)\simeq 1-\frac{Mr^2}{3\sqrt{\pi}\theta^{\frac{3}{2}}}.
\label{ref3}
\end{eqnarray}
Therefore in this limit the metric (\ref{1.04}) reduces to a de-Sitter metric with cosmological constant 
\begin{eqnarray}
\Lambda_\theta=\frac{M}{3\sqrt\pi\theta^{3/2}}
\label{ref4}
\end{eqnarray}
which has a constant scalar curvature, given by, 
\begin{eqnarray}
R_\theta=\frac{4M}{\sqrt\pi\theta^{3/2}}.
\label{ref1}
\end{eqnarray}
Consequently there is no curvature singularity present any more, instead one finds a de-Sitter core of constant positive curvature surrounding the close vicinity of the singularity at $r=0$. This is in agreement with \cite{Frolov,Dym} where the existence of the inner de-Sitter core was mentioned. Taking the commutative limit $\theta\rightarrow 0$ in (\ref{ref1}) immediately manifests the singularity.

 It is interesting to note that the noncommutative metric (\ref{1.04}) is still stationary, static and spherically symmetric as in the commutative case. One or more of these properties is usually violated for other approaches \cite{Obre,Chai,Saha,Kobak} of introducing noncommutativity, particularly those based on Seiberg-Witten maps that relate commutative spaces with noncommutative ones.

      The event horizon of the black hole can be found by setting ${(g_\theta)}_{tt}\Big|_{r=r_h}=0$ in (\ref{1.04}), which yields,
\begin{eqnarray}
r_h=\frac{4M}{\sqrt\pi}\gamma(\frac{3}{2},\frac{r^2_h}{4\theta}).
\label{1.05}
\end{eqnarray}
Keeping upto the leading order $\frac{1}{\sqrt{\theta}}e^{-{M^2}/{\theta}}$, we find
\begin{eqnarray}
r_h \simeq 2M\left(1-\frac{2M}{\sqrt{\pi\theta}}e^{{-M^2}/{\theta}}\right)  
\label{1.06}
\end{eqnarray}

       Now for a general stationary, static and spherically symmetric space time the Hawking temperature ($T_h$) is related to the surface gravity ($\kappa$) by the following relation \cite{Majhi1}
\begin{eqnarray}
T_h=\frac{\kappa}{2\pi};\,\,\,\kappa = [\frac{1}{2}\frac{d{(g_\theta)}_{tt}}{dr}]_{r=r_h}. 
\label{1.061}
\end{eqnarray}
Therefore the Hawking temperature for the noncommutative Schwarzschild black hole is given by,
\begin{eqnarray}
T_h = {\frac{1}{4\pi}}\left[{\frac{1}{r_h}}- {\frac{r_h^2}{4\theta^{3/2}}}\frac{e^-{\frac{{r_h}^2}{4\theta}}}{\gamma({\frac{3}{2}},{\frac{r^2_h}{4\theta}})}\right].
\label{1.08}
\end{eqnarray}
 To write the Hawking temperature in the regime $\frac{r^2_h}{4\theta}>>1$ as a function of $M$ we will use (\ref{1.06}). Keeping upto the leading order in $\theta$ we get
\begin{eqnarray}
T_h\simeq\frac{1}{8{\pi}M}\left[1-\frac{4M^3}{{\sqrt\pi} {\theta}^{3/2}}{e^{-M^2/\theta}}\right].
\label{1.10}
\end{eqnarray}
We will now use the Gibbs form of first law of thermodynamics,  
\begin{eqnarray}
dS_{{\textrm{bh}}}=\frac{dM}{T_h}.
\label{1.1}
\end{eqnarray}
to calculate the Bekenstein-Hawking entropy. Upto the leading order in $\theta$ this is given by \cite{Majhi1},
\begin{eqnarray}
S_{\textrm{bh}}=\frac{A}{4}; \,\,\ A=4\pi r_h^2\simeq 16\pi M^2-64\sqrt{\frac{\pi}{\theta}}M^3e^{-\frac{M^2}{\theta}}.
\label{1.13}
\end{eqnarray}
This is functionally identical to the Benkenstein-Hawking area law in the commutative space.

\begin{figure}[h] 
\centering
\includegraphics[angle=0,width=7cm,keepaspectratio]{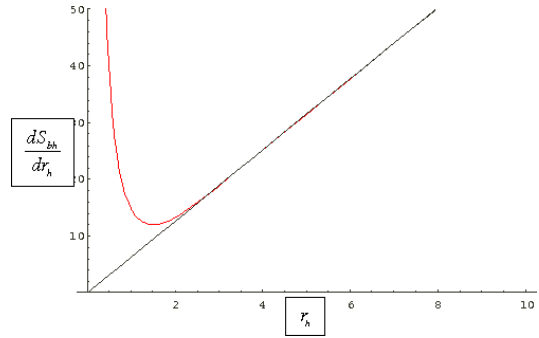}
\caption[]{\it $\frac{dS_{\textrm{bh}}}{dr_h}$ Vs. $r_h$ plot. $\frac{dS_{\textrm{bh}}}{dr_h}$ is plotted in units of $4\theta$ and $r_h$ is plotted in units of $2\sqrt\theta$. Red curve: for eq. (\ref{1.3}), Black curve: for eq. (\ref{1.4}).}
\label{fig1}
\end{figure}

      We have analytically seen above that in the limit $\frac{r^2_h}{4\theta}>>1$ the noncommutative version of the semiclassical Bekenstein-Hawking area law holds upto the leading order in $\theta$. This motivates us to see whether this law holds for all orders in $\theta$, irrespective of the limit we have mentioned. Since analytically it seems very difficult, this issue will be discussed by a graphical analysis. It will be always useful for us to write the right hand side of (\ref{1.1}) in terms of the horizon $r_h$ of the black hole. Using (\ref{1.05}) we have
\begin{eqnarray}
dM= \frac{\sqrt{\pi}}{4\gamma(\frac{3}{2},\frac{r^2_h}{4\theta})}\Big[1-\frac{r^3_h}{4{\theta}^{\frac{3}{2}}}\frac{e^{-\frac{r^2_h}{4\theta}}}{\gamma(\frac{3}{2},\frac{r^2_h}{4\theta})}\Big]dr_h.
\label{1.2}
\end{eqnarray}
Substituting this and (\ref{1.08}) in (\ref{1.1}), we get a closed form relation,
\begin{eqnarray}
\frac{dS_{{\textrm{bh}}}}{dr_h}=\frac{\pi^{\frac{3}{2}}r_h}{\gamma(\frac{3}{2},\frac{r^2_h}{4\theta})}.
\label{1.3}
\end{eqnarray}
This will be compared graphically with the quantity $\frac{dS_{{\textrm{bh}}}}{dr_h}$ calculated from the semiclassical Bekenstein-Hawking area law (\ref{1.13}). This yields, using (\ref{1.13}),
\begin{eqnarray}
\frac{dS_{{\textrm{bh}}}}{dr_h}\Big|_{\textrm{semiclassical}}=2\pi r_h.
\label{1.4}
\end{eqnarray}
Now $\frac{dS_{{\textrm{bh}}}}{dr_h}$ is plotted as a function of $r_h$ (for both equations (\ref{1.3}) and (\ref{1.4})) in figure (\ref{fig1}). 
It is interesting to see that semiclassical area law still holds for $r_h\gtrsim4.8\sqrt\theta$ since the two curves exactly coincide. To further understand this issue we solve for $r_h$ by equating (\ref{1.3}) and (\ref{1.4}) to obtain,
\begin{eqnarray}
\gamma\Big(\frac{3}{2},\frac{r_h^2}{4\theta}\Big)=\frac{\sqrt{\pi}}{2}
\label{gamma}
\end{eqnarray}
which is put in the form,
\begin{eqnarray}
\int_0^{\frac{r_h^2}{4\theta}}dt \sqrt{t} e^{-t}=\frac{\sqrt{\pi}}{2}
\label{int}
\end{eqnarray}
A numerical analysis yields the saturated bound for $\frac{r_h^2}{\theta}$ as $(4.8)^{2}$. This shows that the linear area law holds for values of $r_h\gtrsim 4.8\sqrt\theta$.

     Now in the region $r_h<4.8\sqrt\theta$ the two curves do not coincide. So there is a deviation from the usual area law. Also one can see from figure (\ref{fig1}) that the curve for (\ref{1.3}) (red curve) attains a minimum value at $r_h=3.0\sqrt\theta$ and then sharply diverges for $r_h<3.0\sqrt\theta$ which is physically unreasonable since the change of Bekenstein-Hawking entropy with the horizon is expected to be unidirectional. This point will be cleared in the following analysis.

\begin{figure}[h] 
\centering
\includegraphics[angle=0,width=7cm,keepaspectratio]{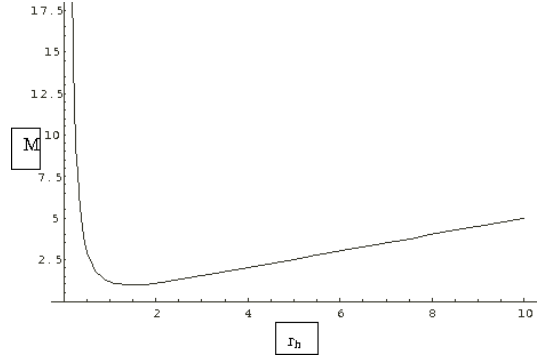}
\caption[]{{\it $M$ Vs. $r_h$ plot. $M$ is plotted in units of $2\sqrt\theta$ and $r_h$ is plotted in units of $2\sqrt\theta$ for eq. (\ref{1.05}).}}
\label{fig2}
\end{figure}

\begin{figure}[h] 
\centering
\includegraphics[angle=0,width=7cm,keepaspectratio]{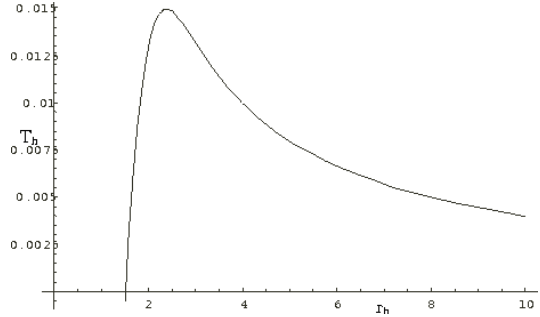}
\caption[]{{\it $T_h$ Vs. $r_h$ plot. $T_h$ is plotted in units of $\frac{1}{\sqrt\theta}$ and $r_h$ is plotted in units of $2\sqrt\theta$ for eq. (\ref{1.08}).}}
\label{fig3}
\end{figure}

       In figure (\ref{fig2}) we plot the black hole mass $M$ as a function of $r_h$ (for equation (\ref{1.05})). It shows there is a minimum value ($M_0=1.9\sqrt\theta$) of $M$ at $r_h=3.0\sqrt\theta$ and noncommutativity introduces new behavior with respect to the standard Schwarzschild black hole \cite{Smail,Park}:\\
(i) Two distinct horizons occur for $M>M_0$: one inner (Cauchy) horizon and one outer (event) horizon. \\
(ii) One degenerate horizon occurs at $r_h=3.0\sqrt\theta$ for $M=M_0$.\\
(iii) No horizon occurs for $M<M_0$.\\  
In the case of $M>>M_0$, the inner horizon shrinks to zero while the outer horizon approaches the Schwarzschild radius $2M$. These features were explained rigorously in \cite{Smail,Park}. Now we plot $T_h$ as a function of $r_h$ in figure (\ref{fig3}) (for equation (\ref{1.08})). It is observed that for $r_h<3.0\sqrt\theta$ there is no black hole because physically $T_h$ cannot be negative. This was explained earlier in \cite{Smail,Park,Majhi1}. Therefore the black hole only exists in the region $r_h\geq3.0\sqrt\theta$ for which there is only one horizon.

       Now the minimum of $\frac{dS_{\textrm{bh}}}{dr_h}$ (see figure (\ref{fig1})) occurs for $r_h=3\sqrt\theta$ which just saturates the limit of physical validity of the black hole. Thus the sharp increase of $\frac{dS_{\textrm{bh}}}{dr_h}$ for $r_h<3\sqrt\theta$ is in the unphysical domain and hence ignored.

        So it is clear from figure (\ref{fig1}) that the semiclassical area law is not satisfied in the region $3.0\sqrt\theta\leq r_h<4.8\sqrt\theta$ while for $r_h\gtrsim4.8\sqrt\theta$ the Bekenstein-Hawking area law holds perfectly. We now find the correction to this area law such that it will describe the entropy for the complete physical region.

\begin{figure}[h] 
\centering
\includegraphics[angle=0,width=7cm,keepaspectratio]{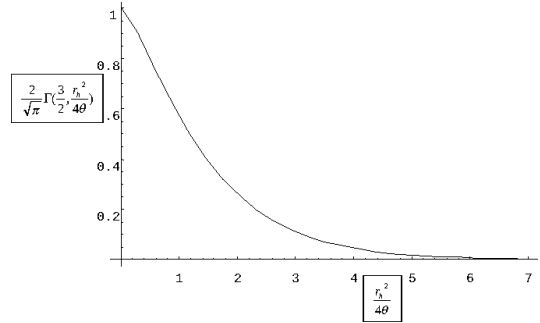}
\caption[]{{\it{$\frac{2}{\sqrt\pi}\Gamma(\frac{3}{2},\frac{r_h^2}{4\theta})$ Vs. $\frac{r_h^2}{4\theta}$ plot.}}}
\label{fig9}
\end{figure}

       To do this we proceed as follows. The first step is to expand (\ref{1.3}) in powers of the upper incomplete gamma function $\Gamma(\frac{3}{2},\frac{r^2_h}{4\theta})$,
\begin{eqnarray}
\Gamma(a,x)=\int_x^\infty t^{a-1}e^{-t}dt
\end{eqnarray}
so that, 
\begin{eqnarray}
\frac{dS_{{\textrm{bh}}}}{dr_h}&=&\frac{\pi^{\frac{3}{2}}r_h}{\frac{\sqrt\pi}{2}-\Gamma(\frac{3}{2},\frac{r^2_h}{4\theta})}
\nonumber
\\
&=&2\pi r_h\Big[1-\frac{2}{\sqrt\pi}\Gamma(\frac{3}{2},\frac{r^2_h}{4\theta})\Big]^{-1}
\nonumber
\\
&=&2\pi r_h\Big[1+\frac{2}{\sqrt\pi}\Gamma(\frac{3}{2},\frac{r^2_h}{4\theta})+\frac{4}{\pi}\Gamma^2(\frac{3}{2},\frac{r^2_h}{4\theta})+\frac{8}{\pi^{\frac{3}{2}}}\Gamma^3(\frac{3}{2},\frac{r^2_h}{4\theta})+........\Big].
\label{2.2}
\end{eqnarray}
The above expansion is valid only when $|\frac{2}{\sqrt\pi}\Gamma(\frac{3}{2},\frac{r_h^2}{4\theta})|<1$. This is proved by a graphical analysis. The plot (\ref{fig9}) shows that $|\frac{2}{\sqrt\pi}\Gamma(\frac{3}{2},\frac{r_h^2}{4\theta})|$ is always less than $1$ for the entire black hole region.

      The first term in (\ref{2.2}) corresponds to the usual area law. The other terms are therefore interpreted as corrections to the area law. To justify this we will take the help of graphical analysis. Taking only the first order correction, $\frac{dS_{{\textrm{bh}}}}{dr_h}$ is written as 
\begin{eqnarray}
{\frac{dS_{{\textrm{bh}}}}{dr_h}}^{(1)}= 2\pi r_h\Big[1+\frac{2}{\sqrt\pi}\Gamma(\frac{3}{2},\frac{r^2_h}{4\theta})\Big].
\label{2.3}
\end{eqnarray}
\begin{figure}[h] 
\centering
\includegraphics[angle=0,width=7cm,keepaspectratio]{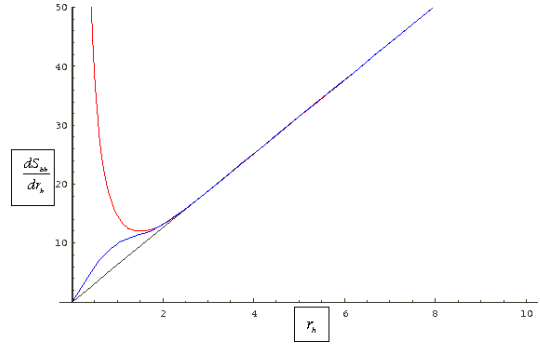}
\caption[]{{\it $\frac{dS_{\textrm{bh}}}{dr_h}$ Vs. $r_h$ plot. {$\frac{dS_{\textrm{bh}}}{dr_h}$ is plotted in units of $4\theta$ and $r_h$ is plotted in units of $2\sqrt\theta$. Red curve: for eq. (\ref{1.3}), black curve: for eq. (\ref{1.4}) and blue curve: for eq. (\ref{2.3})}}.}
\label{fig5}
\end{figure}
The variation of $\frac{dS_{{\textrm{bh}}}}{dr_h}$ versus $r_h$ for equations (\ref{1.3}), (\ref{1.4}) and (\ref{2.3}) is shown in figure (\ref{fig5}). It is observed that the blue curve (corresponding to (\ref{2.3})) has the correct linear behaviour for $r_h\gtrsim4.8\sqrt\theta$. Below this it agrees with the red curve almost till the extremal (physical) limit $r_h=3.0\sqrt\theta$, near which it shows a slight deviation. To improve this situation, the next order correction in (\ref{2.2}) is included,
\begin{eqnarray}
{\frac{dS_{{\textrm{bh}}}}{dr_h}}^{(2)}= 2\pi r_h\Big[1+\frac{2}{\sqrt\pi}\Gamma(\frac{3}{2},\frac{r^2_h}{4\theta})+\frac{4}{\pi}\Gamma^2(\frac{3}{2},\frac{r^2_h}{4\theta})\Big].
\label{2.4}
\end{eqnarray}
\begin{figure}[h] 
\centering
\includegraphics[angle=0,width=7cm,keepaspectratio]{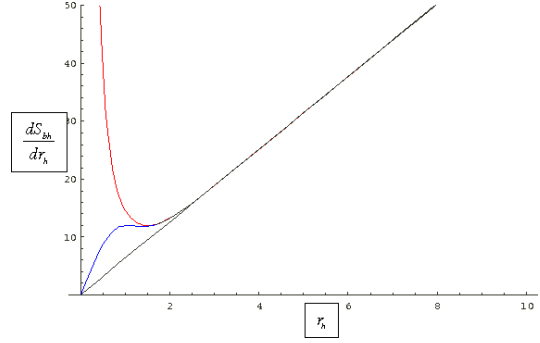}
\caption[]{{\it{$\frac{dS_{\textrm{bh}}}{dr_h}$ Vs. $r_h$ plot. $\frac{dS_{\textrm{bh}}}{dr_h}$ is plotted in units of $4\theta$ and $r_h$ is plotted in units of $2\sqrt\theta$. Red curve: for eq. (\ref{1.3}), black curve: for eq. (\ref{1.4}) and blue curve: for eq. (\ref{2.4})}}.}
\label{fig6}
\end{figure}
This is now plotted in figure (\ref{fig6}) along with equations (\ref{1.3}) and (\ref{1.4}). It shows that the blue curve coincides with the red curve for the entire physical domain $r_h\geq3.0\sqrt\theta$. Incidentally, if the third order correction had been included (see figure (\ref{fig7})), the matching would extend below the extremal limit $r_h=3.0\sqrt\theta$. In fact the curves begin to coincide from $r_h=2.6\sqrt\theta$ which actually lies in the unphysical domain and hence is inconsequential. Therefore we conclude that it is both necessary and sufficient to take upto the second order correction in the variation of the Bekenstein-Hawking entropy with the horizon $r_h$ of the black hole and (\ref{2.4}) should eventually lead to the required correction to the area law in the region of our interest. Now integrating over $r_h$, (\ref{2.4}) yields
\begin{eqnarray}
S_{{\textrm{bh}}}&=& \pi r^2_h-\sqrt{\frac{\pi}{\theta}}~r^3_h e^{-\frac{r^2_h}{4\theta}}-6\sqrt{\pi\theta}~r_h e^{-\frac{r^2_h}{4\theta}}-6\pi\theta\Big(1-{\textrm {Erf}}(\frac{r_h}{2\sqrt\theta})\Big)
\nonumber
\\
&+&2\sqrt\pi ~r^2_h \Gamma(\frac{3}{2},\frac{r^2_h}{4\theta})+8\int r_h\Gamma^2(\frac{3}{2},\frac{r^2_h}{4\theta})dr_h. 
\label{2.5}
\end{eqnarray}
This is the desired expression for the entropy in the entire physical region of the black hole that is valid to all orders in $\theta$. Taking the large radius limit ($\frac{r_h^2}{4\theta}>>1$) and keeping terms upto the leading order ($\frac{1}{\sqrt\theta}e^{-\frac{1}{\theta}}$) immediately reproduces (\ref{1.13}).

        Expressing (\ref{2.5}) in terms of the semiclassical noncommutative area $A=4\pi r_h^2$ (\ref{1.13}) the cherished area law is obtained, 
\begin{eqnarray}
S_{{\textrm{bh}}}&=& \frac{A}{4}-\frac{A^{\frac{3}{2}}}{8\pi\sqrt\theta} e^{-\frac{A}{16\pi\theta}}-3\sqrt{\theta A} e^{-\frac{A}{16\pi\theta}}-6\pi\theta\Big(1-{\textrm {Erf}}(\frac{1}{4}\sqrt\frac{A}{\pi\theta})\Big)
\nonumber
\\
&+&\frac{A}{2\sqrt\pi} \Gamma(\frac{3}{2},\frac{A}{16\pi\theta})+\frac{1}{\pi}\int \Gamma^2(\frac{3}{2},\frac{A}{16\pi\theta})dA. 
\label{2.51}
\end{eqnarray}
\begin{figure}[h] 
\centering
\includegraphics[angle=0,width=7cm,keepaspectratio]{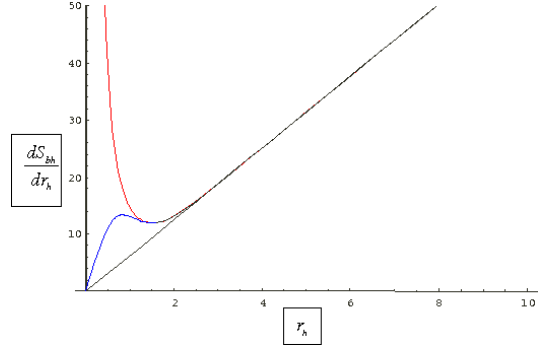}
\caption[]{{\it{$\frac{dS_{\textrm{bh}}}{dr_h}$ Vs. $r_h$ plot. $\frac{dS_{\textrm{bh}}}{dr_h}$ is plotted in units of $4\theta$ and $r_h$ is plotted in units of $2\sqrt\theta$. Red curve: for eq. (\ref{1.3}), black curve: for eq. (\ref{1.4}) and blue curve: for third order correction.}}}
\label{fig7}
\end{figure}
The first term yields the noncommutative version of the famous Bekenstein-Hawking semiclassical area law. The other terms are the corrections to the area law. These corrections, contrary to naive expectations \cite{Majhi2}, do not involve any logarithmic terms. Rather, they involve exponentials of the noncommutative semiclassical area $A$ as well as the error function. Taking the large area limit $(\frac{A}{16\pi\theta}>>1)$ and retaining terms upto the leading order $(\frac{1}{\sqrt\theta}e^{-\frac{1}{\theta}})$, the general structure in (\ref{2.51}) reduces to (\ref{1.13}). Finally, in the commutative limit $\theta\rightarrow 0$, all terms except the $\frac{A}{4}$ term separately vanish and the usual semiclassical Bekenstein-Hawking area law is reproduced.

    There is a further point that deserves some attention. From (\ref{1.01}) it is observed that $r_\theta=2\sqrt\theta$ might be interpreted as the radius of some sphere where noncommutative effects cannot be ignored. In that case the particular combination ($\frac{A}{16\pi\theta}$) appearing in (\ref{2.51}) could be regarded as the ratio between the areas of the black hole and the noncommutative sphere.

       A simple physical consistency check is now done. The point is that at the extremal limit $r_h=3.0\sqrt\theta$ (which corresponds to the zero temperature degenerate horizon state) the entropy must vanish. To show this the appropriate limit of (\ref{2.5}) from $r_h=3.0\sqrt\theta$ upto some arbitrary $r_h$ is taken,

\begin{eqnarray}
S_{{\textrm{bh}}}\Big|_{r_h=3.0\sqrt\theta}^{r_h}&=& \Big[\pi r^2_h-\sqrt{\frac{\pi}{\theta}}~r^3_h e^{-\frac{r^2_h}{4\theta}}-6\sqrt{\pi\theta}~r_h e^{-\frac{r^2_h}{4\theta}}-6\pi\theta\Big(1-{\textrm {Erf}}(\frac{r_h}{2\sqrt\theta})\Big)
\nonumber
\\
&+&2\sqrt\pi ~r^2_h \Gamma(\frac{3}{2},\frac{r^2_h}{4\theta})+8\int r_h\Gamma^2(\frac{3}{2},\frac{r^2_h}{4\theta})dr_h\Big]_{r_h=3.0\sqrt\theta}^{r_h}. 
\label{2.52}
\end{eqnarray}
\begin{figure}[h] 
\centering
\includegraphics[angle=0,width=7cm,keepaspectratio]{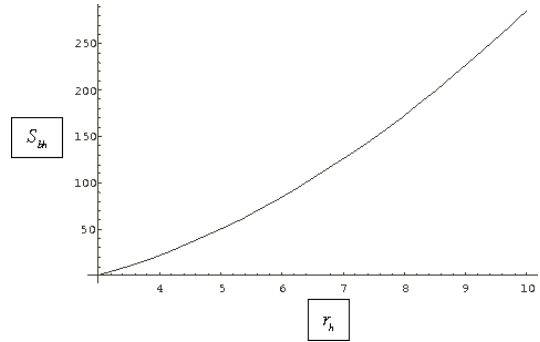}
\caption[]{{\it{$S_{\textrm{bh}}$ Vs. $r_h$ plot. $S_{\textrm{bh}}$ is plotted in units of $\theta$ and $r_h$ is plotted in units of $\sqrt\theta$.}}}
\label{fig8}
\end{figure}
A numerical plot (figure \ref{fig8}) clearly reveals that $S_{\textrm{bh}}=0$ at the extremal point $r_h=3.0\sqrt\theta$. This shows the consistency of the approach.\\\\\  

   To conclude, we have made a detailed investigation of the Hawking temperature, entropy and the area law for a Schwarzschild black hole whose metric is modified by effects of noncommutative spacetime. The noncommutative version of the semiclassical Bekenstein-Hawking area law (\ref{1.13}) holds in the region $r_h\geq4.8\sqrt\theta$. The linear relation between entropy and area is violated below this horizon radius till the extremal point $r_h=3.0\sqrt\theta$. From a graphical analysis we find the correction terms to the area law (given in (\ref{2.51})) to cover the complete physical domain ($r_h\geq3.0\sqrt\theta$) of the black hole. The correction terms involve exponentials as well as error functions.

   We have also discussed the nontrivial behavior of the noncommutative Schwarzschild metric (\ref{1.04}) at different length scales. In the large $r$ limit ($r>>\sqrt\theta$) outside the horizon the metric (\ref{1.04}) behaves like standard Schwarzschild solution, whereas, in the small $r$ limit ($r<<\sqrt\theta$) inside the horizon we get a de-Sitter metric with constant positive curvature. As a consequence the singularity at $r=0$ is removed for the solution (\ref{1.04}). The stress tensor corresponding to the noncommutative geometry describes a smooth transition from the usual vacuum state at infinity to an anisotropic vacuum state at intermediate ranges finally passing on again to an isotropic state at $r\rightarrow 0$. Furthermore, in the commutative limit ($\theta\rightarrow 0$) the curvature singularity reappears as one can expect from the behaviour of standard Schwarzshild metric.\\\\ 

{\bf{Acknowledgment}}\\
One of the authors (SKM) thanks the Council of Scientific and Industrial Research (CSIR), Government of India, for financial support.

\end{document}